\documentclass[a4paper, amsfonts, amssymb, amsmath, reprint, showkeys, nofootinbib,superscriptaddress]{revtex4-2}
\usepackage{amsmath,empheq}
\usepackage{braket}
\usepackage{graphicx}
\usepackage{textcomp}
\usepackage{gensymb}
\usepackage{hyperref}
\usepackage{mathtools}
\usepackage{dsfont}
\usepackage{bbold}

\begin{document}
\title{Instabilities near ultrastrong coupling in microwave optomechanical cavity}

\author{Soumya~Ranjan~Das}
\author{Sourav~Majumder}
\author{Sudhir~Kumar~Sahu}
\author{Ujjawal~Singhal}
\author{Tanmoy~Bera}
\author{Vibhor~Singh}
\email[]{v.singh@iisc.ac.in} 
\affiliation{Department of Physics, Indian Institute of Science, 
	Bangalore-560012 (India)}

\begin{abstract}
With artificially engineered systems, it is now possible 
to realize the coherent interaction rate, which can become 
comparable to the mode frequencies, a regime known as 
ultrastrong coupling (USC).
We experimentally realize a cavity-electromechanical device
using a superconducting waveguide cavity and a  
mechanical resonator.
In the presence of a strong pump, the
mechanical-polaritons splitting can nearly 
reach 81\% of the mechanical frequency, overwhelming
all the dissipation rates.
Approaching the USC limit, the steady-state response becomes unstable. 
We systematically measure the boundary of 
the unstable response while varying the pump 
parameters.
The unstable dynamics display rich phases, such as 
self-induced oscillations, period-doubling bifurcation, 
period-tripling oscillations, and ultimately leading to 
the chaotic behavior. 
The experimental results and their theoretical
modeling suggest the importance of residual 
nonlinear interaction terms in the weak-dissipative 
regime.

\end{abstract}
	
\date{\today}
	
	
	\maketitle
	
	\section{Introduction:}\label{intro}
	Radiation-pressure interaction is fundamental to 
	the cavity-optomechanical systems consisting 
	of  a mechanical mode coupled to an electromagnetic 
	mode (EM) \cite{aspelmeyer_cavity_2014}. 
	With technological advancements, 
	cavity optomechanical devices have been
	successful in controlling the low-frequency mechanical 
	mode down to their quantum 
	regime \cite{barzanjeh_optomechanics_2022}. 
	Several demonstrations pertaining to the quantum state 
	preparation \cite{teufel_sideband_2011,chan_laser_2011,
		wollack_quantum_2022,bild_schrodinger_2022} and 
	entanglement \cite{palomaki_coherent_2013,riedinger_remote_2018,ockeloen-korppi_stabilized_2018,kotler_direct_2021}, 
	signal transduction \cite{andrews_bidirectional_2014,forsch_microwave--optics_2020,mirhosseini_superconducting_2020}, 
	and topological physics using the mechanical modes 
	have been shown \cite{cha_experimental_2018,youssefi_topological_2022}.

\begin{figure}
		\centering
		\includegraphics[width = 75 mm]{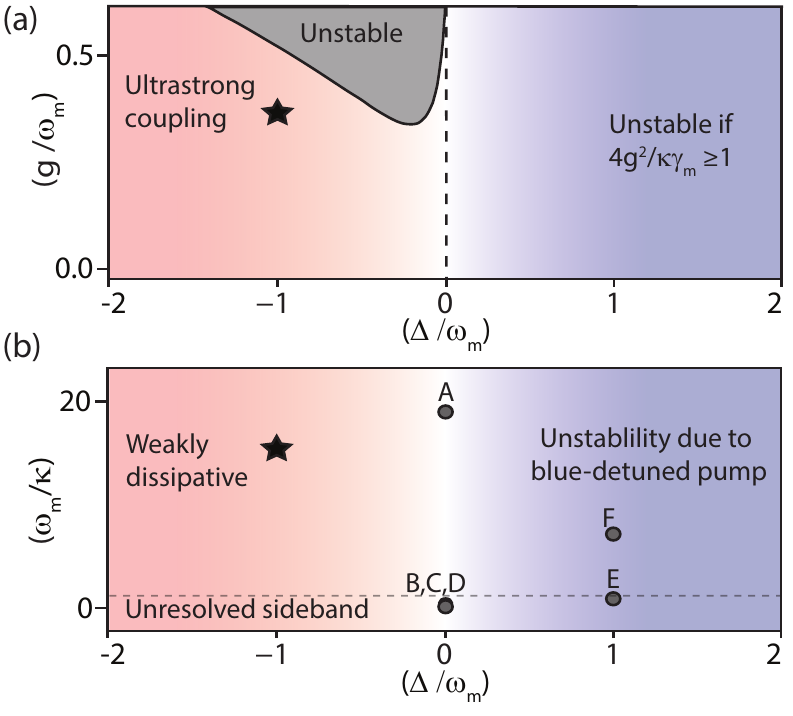}
		\caption{(a) The schematic of the 
			parameter space of a cavity-electromechanical system in the steady state marking the USC regime and the region of unstable response. This study has been marked by $\star$.
			(b) comparison the sideband-resolution 
			parameter of this study and 
			the earlier studies on instabilities. 
			The data points A to F are from 
			refs \cite{carmon_chaotic_2007, navarro-urrios_nonlinear_2017, leijssen_nonlinear_2017, buters_experimental_2015, monifi_optomechanically_2016, shin_-chip_2022}, respectively. The symbols $g$, $\omega_m$, $\kappa$, and  $\Delta$ represent 
			the optomechanical coupling strength, the mechanical frequency, 
			cavity dissipation rate, and the pump detuning from the 
			cavity resonant frequency, respectively.}
		\label{fig0}
	\end{figure}

	The coherent coupling rate, characterizing the 
	interaction between the EM mode $(\omega_c)$ and 
	the mechanical mode $(\omega_m)$, is a key 
	figure of merit in such 
	devices \cite{aspelmeyer_cavity_2014,barzanjeh_optomechanics_2022}. 
	The energy dissipation rates of the
	two modes $(\kappa$, $\gamma_m)$ capture the 
	incoherent coupling with their thermal baths. 
	Based on the relative strengths of these rates, 
	several interesting scenarios are 
	feasible.
	When the coherent coupling rate ($g$) 
	exceeds the dissipative coupling rates of the 
	two modes $(g\gg\kappa,\gamma_m)$, the two modes 
	hybridize, resulting in new 
	eigenstates \cite{teufel_circuit_2011,verhagen_quantum-coherent_2012}.
	Further, when the coherent coupling rate becomes
	a significant fraction of the mode frequencies, 
	the composite system enters the
	``ultrastrong coupling" (USC) 
	limit \cite{ciuti_input-output_2006}.
	In this limit, the two modes hybridize in a non-trivial 
	way leading to an entangled ground state in the 
	quantum regime \cite{ciuti_quantum_2005, hofer_quantum_2011}.
	The USC limit has been experimentally demonstrated
	in several systems where two modes interact
	nearly resonantly \cite{forn-diaz_ultrastrong_2019,kockum_ultrastrong_2019}.
	In cavity optomechanical systems, however, the 
	EM mode and mechanical mode interact dispersively
	$(\omega_c\gg\omega_m)$.
	The nonlinear radiation-pressure interaction  
	can be described by $H_{i}/\hbar = -{g_0}a^{\dagger}a(b+b^{\dagger})$, 
	where $g_0$ is the single-photon coupling strength 
	and $a(b)$'s are the ladder operators for
	the cavity(mechanical) mode.
	In the presence of a strong coherent pump, 
	the interaction Hamiltonian can be linearized to 
	$H_{i}/\hbar\simeq -g(a+a^{\dagger})(b+b^{\dagger})$,
	where $g=g_0\sqrt{n_d}$ is the parametric coupling rate and
	$n_d$ is the number of the pump photons
	in the cavity.
	With the ability to control the parametric coupling rate,
	several regimes, such as quantum coherent
	coupling, and steady-state quantum entanglement between
	the two modes can be reached \cite{verhagen_quantum-coherent_2012,hofer_entanglement-enhanced_2015,kotler_direct_2021}. 
	Ultimately, owing to 
	the nonlinear nature of the radiation-pressure interaction, 
	the response becomes unstable, as shown schematically
	in Fig.~\ref{fig0}.
	Indeed, various phenomena in the unstable region such as limit cycle, period 
	doubling bifurcations, and chaos have been extensively 
	studied \cite{marquardt_dynamical_2006,lorch_laser_2014,bakemeier_route_2015,schulz_optomechanical_2016,djorwe_frequency_2018}. 
	Experimentally, these effects have been primarily
	explored in the strong dissipative regime $(\omega_m\lesssim\kappa)$ 
	or with the blue-detuned pump \cite{kippenberg_analysis_2005,carmon_chaotic_2007,buters_experimental_2015,monifi_optomechanically_2016,navarro-urrios_nonlinear_2017,leijssen_nonlinear_2017,shin_-chip_2022} (see Fig.~\ref{fig0}).
	The instabilities near the ultrastrong 
	coupling limit, however, allow to explore
	the nonlinear dynamics of the cavity optomechanical 
	system in the \textit{weakly dissipative }
	limit ($\gamma_m, \kappa\ll 2g\lesssim\omega_m$).
	The nonlinear dynamics with weak dissipation is unique
	and is predicted to show transient chaos, quasi-periodic 
	route to chaos and lower threshold powers for the onset 
	of chaos \cite{miri_optomechanical_2018,roque_nonlinear_2020}.
	%
	
	%
	Here, we use a cavity electromechanical device in the 
	microwave domain to probe the route to chaos when it 
	is operated into the USC limit.
	We first demonstrate the USC by performing the
	spectroscopic and time-domain measurements.
	We probe the stability of the device when the pump
	detuning near the red sideband and injected power are varied. 
	The unstable region shows very rich phases in the 
	parameter space, such as the self-induced oscillation, 
	period-doubling bifurcations, period-tripling 
	oscillations, and chaotic 
	behavior \cite{bakemeier_route_2015,miri_optomechanical_2018,
		roque_nonlinear_2020}.
	We find that the measured threshold 
	powers for the onset of instabilities are lower than the ones predicted from a nonlinear model 
	considering the optomechanical interaction and 
	a Kerr-term in the cavity.

	\section{Experimental details:}
	
	\begin{figure}
		\centering
		\includegraphics[width = 80 mm]{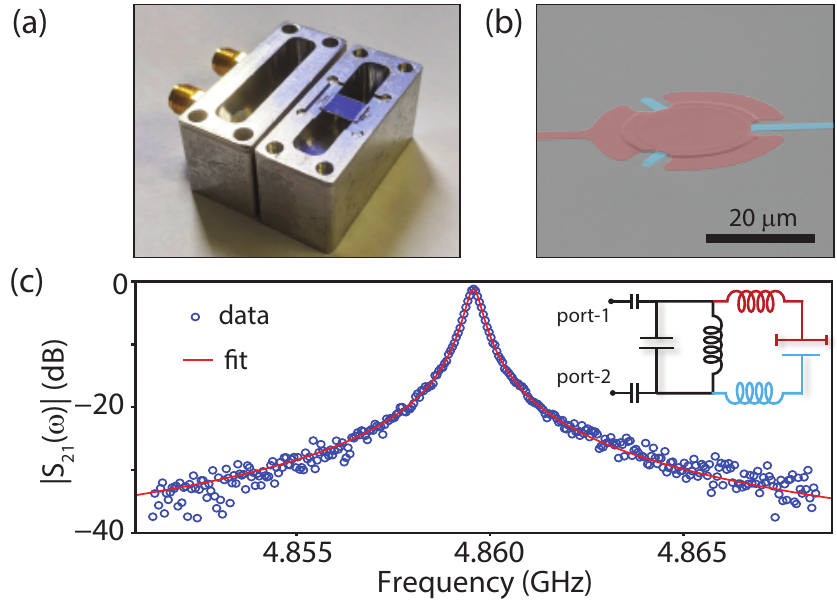}
		\caption{(a) Image of the waveguide cavity along 
			with a patterned substrate.
			The cavity has dimensions of $26\times26\times6$~mm$^3$.
			(b) False color image of the mechanical resonator forming 
			a parallel plate capacitor with another plate on the substrate. The separation 
			between the capacitor plates at room temperature is 
			approximately 200~nm.
			(c) Measurement of the voltage transmission coefficient $|S_{21}|$ of 
			the device at the base temperature. 
			The inset shows the equivalent circuit diagram of the 
			cavity electromechanical device.}
		\label{fig1}
	\end{figure}

	We use the three-dimensional cavity-based 
	platform to realize the cavity-electromechanical 
	device \cite{yuan_large_2015,noguchi_ground_2016,gunupudi_optomechanical_2019}.
	The waveguide cavity-based electromechanical device 
	offers a higher dynamic range for the pumped photons, 
	which is highly desirable to reach the USC limit
	\cite{gunupudi_optomechanical_2019,peterson_ultrastrong_2019}.
	As shown in Fig.~\ref{fig1}(a) and~\ref{fig1}(b), the device consists of 
	a rectangular waveguide cavity, and a drumhead-shaped mechanical resonator in the form of 
	a parallel plate capacitor patterned on a sapphire chip.
	The patterned sapphire chip fabricated with aluminum is 
	placed at the center of the cavity.
	The electrical pads to the drumhead are then 
	directly wire-bonded to the cavity 
	walls to integrate with the cavity 
	mode \cite{gunupudi_optomechanical_2019}. 
	The sample-mounted cavity is cooled down to 20~mK in 
	a dilution fridge. 
	Fig.~\ref{fig1}(c) shows the measurement of the 
	cavity transmission at the base temperature. 
	The bare cavity is designed to have the 
	fundamental resonant mode frequency of $7.5$~GHz. 
	However, the electromechanical 
	capacitor perturbs the mode shape significantly, 
	and lowers the mode frequency to $\omega_c/2\pi\approx4.86$~GHz.
	The reduction in the resonant frequency of the 
	cavity results from the electromechanical 
	capacitance and the inductance of the connecting 
	electrodes introduced after the addition of 
	a patterned sapphire chip.
	We measure the input, output, and 
	the internal dissipation rates of 
	$\kappa_{e1}/2\pi\approx90$~kHz, 
	$\kappa_{e2}/2\pi\approx190$~kHz, and 
	$\kappa_{i}/2\pi\approx100$~kHz,
	respectively.
	At low temperatures, we estimated that 
	the gap between the electromechanical capacitor
	plates reduces to approximately 32~nm due to thermal 
	contraction \cite{wollman_quantum_2015,reed_faithful_2017},
	which helps in achieving a single photon coupling rate 
	$g_0/2\pi$ of 165~Hz.

	\section{Ultrastrong coupling regime}

	\begin{figure}
		\centering
		\includegraphics[width = 80mm]{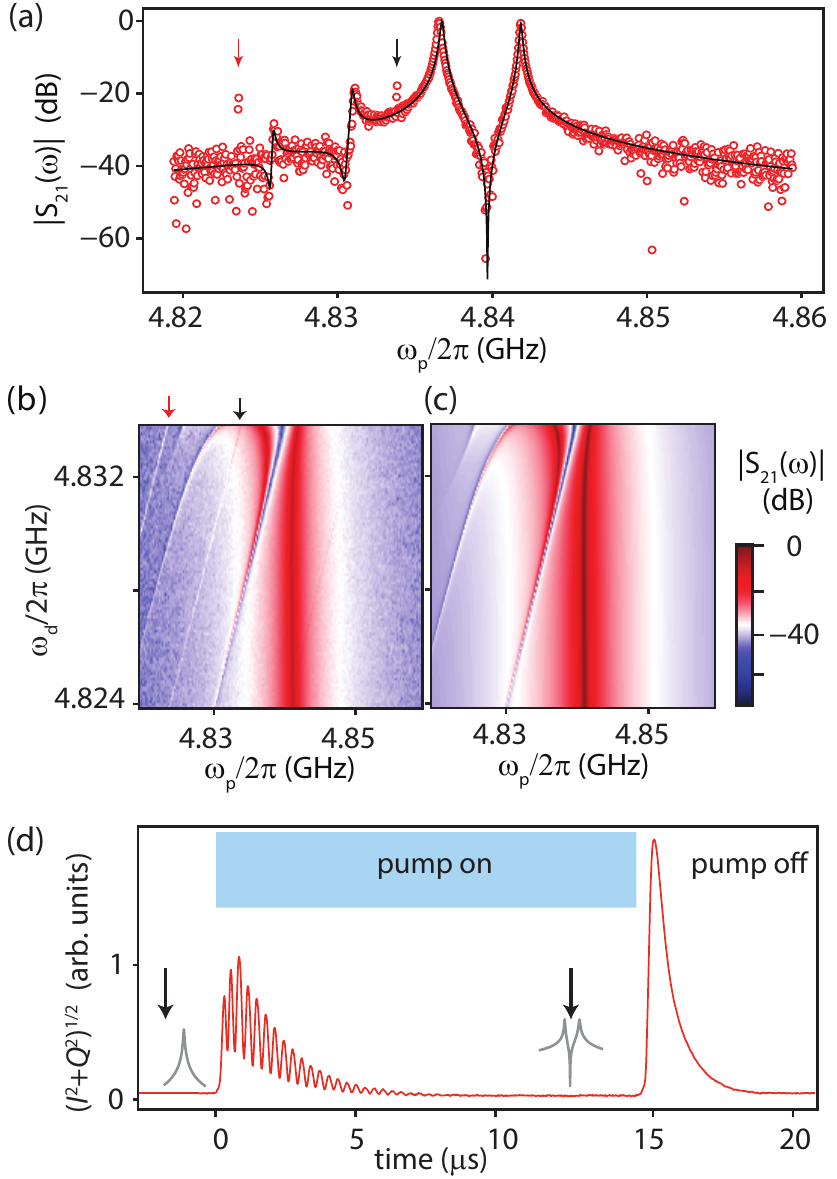}
		\caption{
			(a) The normalized magnitude of the cavity 
			transmission $|S_{21}(\omega)|$ (red-circles) while applying 
			a strong pump near the red sideband. 
			The black line shows the calculated $|S_{21}(\omega)|$ 
			while including the static Kerr shift 
			of the cavity.
			(b) The colorplot of measured $|S_{21}(\omega)|$
			as the pump frequency $\omega_d$ is varied at a fixed 
			pump power of -31 dBm at the cavity. 
			The black arrow shows the position of the pump signal, 
			while the red arrow indicates another weakly coupled 
			mechanical mode. A weak probe signal -88~dBm at the cavity
			is used to generate the colorplot.
			(c) Colorplot of $|S_{21}(\omega)|$ obtained from the 
			calculations. (d) Measurement of the amplitude
			of the transmitted probe signal in the time domain 
			while modulating the interaction strength. 
			The pump frequency is set near the lower 
			mechanical sideband. The position of the probe
			signal is schematically represented by the black arrows
			relative to the steady-state cavity transmission curves (shown in gray).}
		\label{fig2}
	\end{figure}
	
	We measure the transmission coefficient $|S_{21}(\omega)|$ 
	through the cavity using a weak probe tone while 
	injecting a pump detuned near the red sideband
	$\omega_c - \omega_m$. At relatively lower pump powers, the
	optomechanically induced absorption setup allows
	us to determine the mechanical frequency $\omega_m/2\pi\approx$~6.32~MHz
	\cite{weis_optomechanically_2010}. 
	At relatively higher pump powers, the response turns into
	two well-separated peaks confirming the new eigenmodes of 
	the system as shown in Fig.~\ref{fig2}(a).
	The peak separation being $0.81\omega_m$ marks the 
	ultrastrong coupling between the mechanical 
	resonator and the cavity.
	The transmission measurement shows the 
	presence of an additional weakly coupled 
	mechanical mode, indicated by the red arrow. 
	Two more features arising from the 
	interference of the down-scattered pump signal 
	and the probe signal can be seen.
	Figure~\ref{fig2}(b) shows the measurement of $|S_{21}|$
	as the frequency of the pump is varied while maintaining a 
	constant power at the signal generator.
	%

	The presence of a strong intracavity pump field 
	leads to a static shift of the equilibrium position 
	of the mechanical resonator, given by 
	$x_s=(2g_0n_d/\omega_{m})x_{zp}$, where $x_{zp}$ is 
	the zero-point motion of the mechanical resonator.
	The shift in the equilibrium position of the mechanical 
	resonator leads to a Kerr shift of the cavity 
	frequency by $-2g_0^2n_d/\omega_m$. 
	The total shift in the cavity 
	frequency comes from the static nature of the
	radiation-pressure force and nonlinear kinetic 
	inductance of the superconducting aluminum film. 
	We emphasize here that at high pump powers, the Kerr shift 
	of the cavity becomes significant, and it must be
	considered to capture the cavity transmission faithfully. 
	In this case, we found a cavity shift of $\sim$1.76 MHz 
	at the maximum pump power used in the experiment. 
	It corresponds to an optomechanical Kerr coefficient
	of 8.6~mHz/photon and a 
	kinetic inductance Kerr coefficient of approximately 5~mHz/photon 
	at the maximum pump power (see the Supplemental Material\cite{suppli}).

	To theoretically model the cavity transmission, we 
	expand the interaction Hamiltonian $H_i$ around the mean field
	of the pump and obtain the quantum-Langevin equations 
	of motion. 
	Without using the rotating-wave approximation
	and by retaining the static Kerr shift of the cavity 
	frequency, the steady-state response can be obtained
	from the inverse of the mode-coupling matrix (see the Supplemental Material\cite{suppli}). 
	The solid line in Fig.~\ref{fig2}(a) and colorplot in
	Fig.~\ref{fig2}(c) show the calculated 
	transmission coefficient using the experimentally 
	determined device parameters. 
	While in general, additional weakly coupled
	mechanical modes can also be included in the calculations, 
	we neglect them here for simplicity.

	The onset of the strong coupling
	allows for a coherent swap of the excitations between 
	the cavity and the mechanical mode.
	It thus enables the high-speed optomechanical 
	swap gates in the ultrastrong coupling limit. 
	To explore the maximum speed of the optomechanical swap,
	we perform time domain measurements in this limit. 
	We modulate the interaction strength 
	$g(t)$, which is controlled by the amplitude of 
	the pump tone. 
	The transmission through the cavity is monitored by 
	applying a weak continuous probe signal 
	near $\omega_c$. 
	To demodulate the probe signal, we first mix it down
	using an external mixer and then sending it to a high-speed 
	lock-in amplifier to further demodulate the quadratures 
	with a short integration time (100~ns) (see the Supplemental Material\cite{suppli}). 

	Fig.~\ref{fig2}(d) shows the measurement of 
	the magnitude of the demodulated signals $(I(t), Q(t))$
	as the interaction strength $g(t)$ is modulated. 
	For this measurement, the interaction strength is 
	modulated to 1.55~MHz. 
	The probe frequency is detuned from the cavity resonant 
	frequency by $(\omega_{p}-\omega_c)/2\pi=372$~kHz. 
	Therefore, the transmission is small even when the pump tone
	is off $(t<0)$.
	In the steady-state when the pump is turned on ($t\approx10~\mu$s),
	the transmission is low again due to the formation of  
	mechanical-polariton modes. 
	Because of the strong static Kerr shift of the cavity, the probe tone appears near the center of the split peaks, resulting in low transmission.
	When interaction is just switched on, the transient response shows the oscillations arising from the coherent 
	energy exchange between the mechanical and the cavity modes. 
	The oscillation frequency of 3.1~MHz corresponds
	to the characteristic swap time of $160$~ns.
	The amplitude of the oscillations decays at a 
	rate $\simeq~\kappa/4$ set by joint dissipation of the two 
	polaritons. 
	When the pump is turned off, the energy stored 
	in the two polariton modes reemerges near the 
	probe frequency, and the amplitude decays 
	at $\kappa/2$. 
	It is important to remark that as we operate close to the USC limit, 
	the modulated pump signal spectrally overlaps with the probe signal
	and adds a small offset in the measurement.
	Additional datasets are provided in the Supplemental Material\cite{suppli}.

	
	\section{Parametic instabilities near ultrastrong coupling}
	
	
	\begin{figure}
		\centering
		\includegraphics[width = 75 mm]{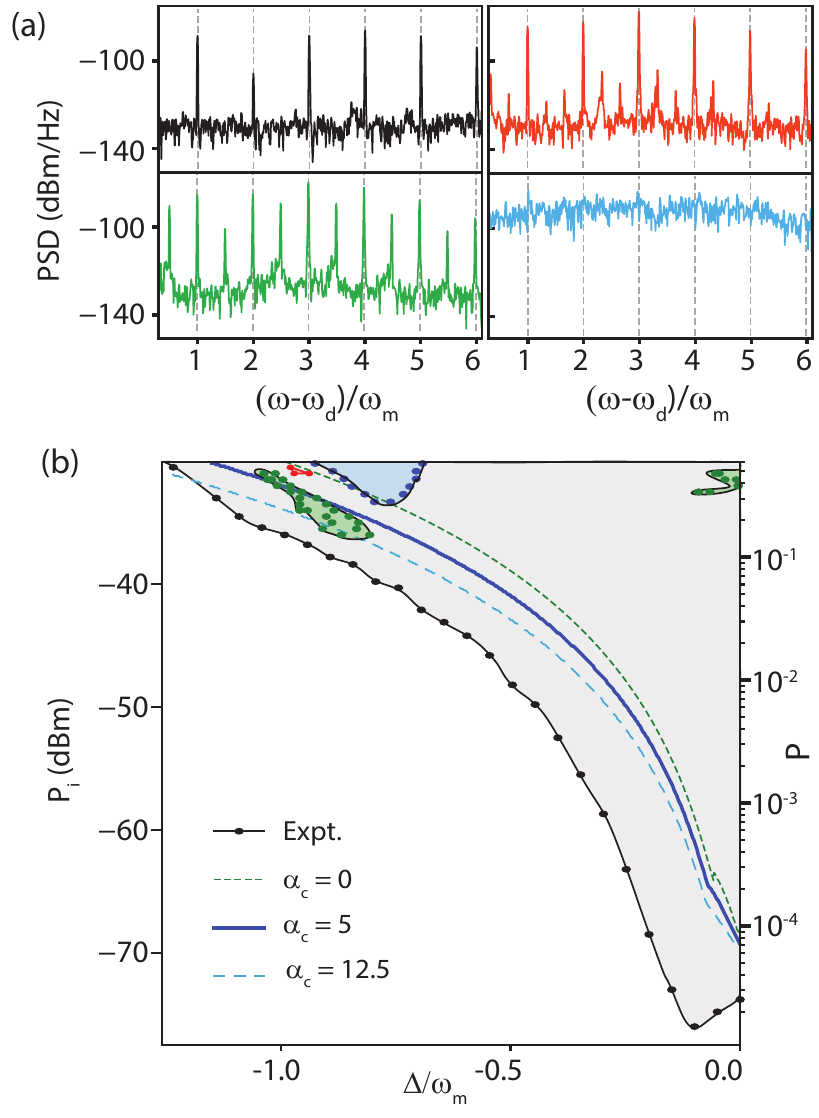}
		\caption{
			(a) Measurement of the single sideband microwave 
			power spectral density (PSD) for different 
			injected pump powers $P_{i}$. 
			Different panels show the self-induced 
			oscillations(black), first period-doubling bifurcations(green), 
			period tripling oscillations(red), and the chaotic behavior(blue).
			(b) The gray region represents the boundary of self-induced 
			oscillations. The black circles are the experimentally 
			measured points.
			At higher injected pump powers,
			the region of period-doubling bifurcations is 
			represented by the green region. The period tripled
			oscillations are shown by the red-colored region. 
			The region of chaotic behavior is shown by the 
			cyan-colored region. Experimentally measured 
			points are shown by the circles of different colors.
			The solid blue line and the dashed lines show the unstable 
			boundary obtained from the theoretical calculation.}
		\label{fig4}
	\end{figure}
	After establishing the USC in the present experiment,
	we now discuss the parametric instabilities arising 
	at the high pump powers. 
	At the core of it, the instabilities stem from the 
	nonlinear interaction between the microwave
	field and the mechanical motion. 
	To experimentally investigate the phase space of the
	parametric instabilities and their nature, we measure 
	the microwave power spectral density (PSD) using a 
	spectrum analyzer while varying the pump power 
	$P_{i}$ and pump detuning $\Delta = (\omega_d-\omega_c)$.
	As $P_{i}$ is increased, the self-induced oscillations 
	appear as multiple peaks separated by $\omega_m$ in 
	the microwave PSD. 
	Fig.~\ref{fig4}(a) shows the PSD of different 
	kinds of responses.
	The top-left panel corresponds to the instability due 
	to the self-induced oscillations where the peaks are 
	separated by $\omega_m$. 
	The bottom-left panel corresponds to the first 
	period-doubling bifurcations (PDB), where 
	the peaks are separated by $\omega_m/2$. 
	The top-right panel shows the period-tripled oscillations
	where the peaks are separated by $\omega_m/3$. 
	The bottom-right panel shows the response where the power 
	is uniformly distributed over a broad range of
	frequencies. It corresponds to the chaotic vibration of the 
	mechanical oscillator undergoing aperiodic oscillations 
	leading to a continuous power spectrum in the output 
	microwave field.
	These different phases of unstable response are
	summarized in Fig.~\ref{fig4}(b).
	The gray-color region represents the parametrically 
	unstable response. The boundary of the gray region
	marks the threshold power for the self-induced 
	oscillations.
	With decrease in $\Delta$, the circulating power 
	in the cavity decreases, and the threshold power 
	for the onset of the instability increases.
	The regions of first PDB, period tripled
	oscillations and chaos are color-coded within the 
	unstable region.
	A discussion on the measurement methodology is
	included in the Supplemental Material\cite{suppli}.

	To understand these results, we use the classical
	nonlinear dynamics approach. We start with the
	full cavity optomechanical Hamiltonian \textit{i.e.}
	$H_i/\hbar = -g_0 \hat{a}^\dagger\hat{a}(\hat{b} + \hat{b}^\dagger)$.
	In addition, motivated by the observation of period-tripling 
	oscillations and the relevance of the kinetic
	inductance at the high pump powers, 
	we include a weak nonlinear term in the cavity Hamiltonian,
	given by $-(\alpha_c/2)(\hat{a}^\dagger\hat{a})^2$.
	Using the semi-classical approximation, we 
	obtain the classical equations of motion (EOM)
	for the cavity and the mechanical quadratures. 
	From EOMs, we find the fixed points and perform
	a linear stability test, which is similar to 
	the Routh-Hurwitz criteria, \textit{i.e.} the solutions
	are stable if and only if all the eigenvalues of the 
	evolution matrix of small perturbations around 
	the fixed points have a negative real part \cite{jeffrey2007}. 
	Calculation details are provided in the Supplemental Material\cite{suppli}.
	Results of these calculations in different limits
	are shown in Fig.~\ref{fig4}(b). 
	We also include a dimensionless 
	power $P=8g_0^2n_0/\omega_m^4$ on the right $y$axis,
	where $n_0$ is defined as the number of photons 
	when the pump is set at the cavity frequency \cite{bakemeier_route_2015,roque_nonlinear_2020}.
	Clearly, the threshold power estimated from the 
	calculations is larger than the one measured in 
	the experiment.
	For comparison, the instability 
	boundary obtained while considering the optomechanical
	Kerr nonlinearity alone $(\alpha_c=0)$, and two
	non-zero values of $\alpha_c/2\pi=5, 12.5$~mHz/photon are also included.  
	We note that even a significantly higher value 
	of $\alpha_c$ does not fully explain the 
	experimental findings suggesting a different origin. 
	Thus, the nonlinearities arising from the 
	optomechanical interaction and the kinetic inductance do not completely capture the threshold for the unstable
	region when a linear stability test is applied.
	In addition, the numerical calculations do not 
	show the period-doubling bifurcations or chaotic
	behavior for the pump parameters used in the
	experiment. 
	In numerical calculations, these effects appear 
	at higher powers than the ones observed in the 
	experiment.
	It thus provides the first experimental evidence 
	that the route to chaos in the USC limit or
	equivalently in the weakly dissipative limit is 
	different from the previously studied cases.
	It suggests that the role of thermal fluctuations, 
	and residual weak nonlinear coupling terms 
	might be relevant in determining the boundary 
	of the unstable region \cite{lemonde_nonlinear_2013,hauer_nonlinear_2023}.
	In particular, during the transitions from 
	self-oscillation to period-doubling oscillation 
	and subsequently to chaotic regions, the mechanical 
	mode remains in a high amplitude state.
	In this case, the role of mechanical Duffing 
	nonlinearity, and resonantly-induced negative
	dissipation might become important \cite{dykman_resonantly_2019}.

	\section{Outlook and conclusion}\label{conclusion}
	
	In conclusion, we have demonstrated the ultrastrong 
	coupling using a superconducting waveguide cavity and 
	a mechanical resonator, where the splitting of the 
	mechanical polaritons becomes nearly 81\% of 
	the mechanical frequency.
	In the time-domain, we measure optomechanical swap
	time of 160~ns, which is nearly 16~times shorter
	than the shortest dissipation time in the device.
	With suitable modification to the thermalization
	of the microwave signals, the cavity can be operated
	in the quantum limit.
	It would enable a wide variety of experiments
	such as the entangled ground state properties 
	of the cavity and the mechanical resonator \cite{kockum_ultrastrong_2019},
	and high speed optomechanical gates \cite{reed_faithful_2017}.
	Using the pump in a pulse mode, the parametric
	coupling can be pushed beyond the USC 
	regime \cite{hofer_quantum_2011,vanner_pulsed_2011}. 
	In addition, the microwave frequency comb 
	generated using the optomechanical-nonlinearity 
	can be a valuable resource for sensing 
	applications \cite{miri_optomechanical_2018}.
	The experiment here, for the first time, explores the 
	unstable response in the steady-state in the weakly 
	dissipative limit \cite{roque_nonlinear_2020}.
	Clearly, the theoretical model based on optomechanical 
	and kinetic inductance nonlinearity  does not 
	account for the lower threshold powers observed in 
	the experiment.
	It thus opens up the possibility of further exploring the role
	of quantum fluctuations \cite{bakemeier_route_2015}, and 
	other weak residual couplings in the interaction 
	Hamiltonian \cite{lemonde_nonlinear_2013}. 
	\begin{acknowledgments}
		The authors thank G.~S.~Agarwal, Manas~Kulkarni, 
		and Tamoghana Ray for their valuable discussions. 
		This material is based upon work supported by 
		the Air Force Office of Scientific Research under 
		award No. FA2386-20-1-4003. 
		V.S. acknowledge the support received under 
		the Core Research Grant by the Department of 
		Science and Technology (India). 
		The authors acknowledge device fabrication facilities 
		at CeNSE, IISc Bangalore, and central facilities 
		at the Department of Physics funded 
		by DST (Govt. of India).
	\end{acknowledgments}
	

%

\end{document}


\title{Supplemental Material: Instabilities near ultrastrong coupling in microwave optomechanical cavity}
	
\author{Soumya~Ranjan~Das}
\author{Sourav~Majumder}
\author{Sudhir~Kumar~Sahu}
\author{Ujjawal~Singhal}
\author{Tanmoy~Bera}
\author{Vibhor~Singh}
\affiliation{Department of Physics, Indian Institute of Science, 
	Bangalore-560012 (India)}

\date{\today}

\maketitle

\section{Device fabrication}

\begin{figure*}
	\begin{center}
		\includegraphics[width = 160 mm]{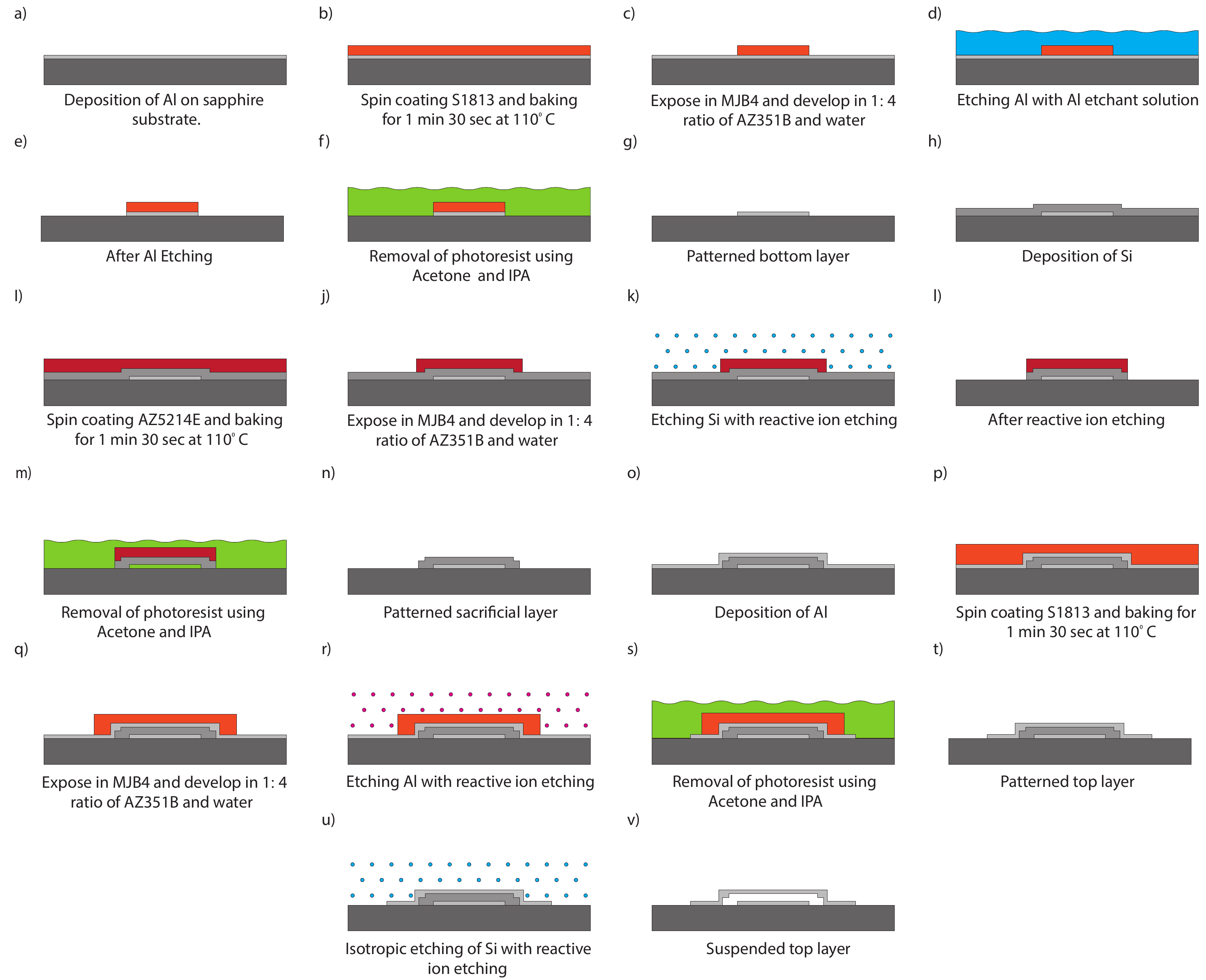}
		\caption{A schematic of the fabrication process summarizing the key steps }
		\label{fabrication_schematic}
	\end{center}
\end{figure*}

The nanofabrication consists of several steps, which are described below. The details of the 
fabrication processes are schematically illustrated 
in the Fig.~\ref{fabrication_schematic}.
We begin with a 2-inch diameter double-side polished 
sapphire wafer and dice it into smaller substrates 
of 5 $\times $ 8 $\text{mm}^{2}$ size. 
Aluminum parallel plate capacitors are then 
fabricated onto these substrates after cleaning.

\textbf{Step-1 Substrate cleaning:}
These diced pieces are heated on a sample holder in 
PG remover for 30~minutes at $80^{\circ}$C and then rinsed with IPA. 
Samples are then dipped in concentrated HNO$_3$ for 10~minutes 
and followed by rinsing with DI water. The samples are then cleaned 
in piranha solution for 10~minutes, followed by rinsing with DI water.

\textbf{Step-2 Patterning of the base electrode:}
Cleaned wafer pieces are loaded in an electron beam evaporator, and then 60~nm thin film of aluminum is deposited with a deposition rate 
of 3~\AA /s. Subsequently, the samples are spin-coated with 
photoresist (S1813). We use a spin speed of 6000~RPM
to minimize the edge bead. The resist-coated samples are baked 
at $110^{\circ}$C for 90~sec. 
The base electrode is patterned using a MJB4 lithography tool, 
followed by development in a 4:1 solution of water and developer 
AZ351B for 6-7~seconds. 
The exposed aluminum is etched with a 3:1 solution of H$_3$PO$_4$ 
and HNO$_3$ and rinsed with DI water. The samples are 
dipped in a 1:2 solution of MF26A and water for 10 seconds 
to remove the aluminum oxide layer, followed by rinsing with 
DI water. Removal of the photoresist is done in 
acetone for 5~minutes, followed by rinsing with IPA. 
Further, the samples are  exposed to O$_2$ plasma with 50~W power 
for 1~minute to remove any residue of photoresist.

\textbf{Step-3 Patterning of the sacrificial layer:}
Silicon of 200~nm is deposited using a sputtering machine 
followed by spin coating with photoresist AZ5214E at 
6000 RPM and followed by baking for 90~seconds at 
$110^{\circ}$C. 
Similar to previous steps, the sacrificial layer is 
patterned with MJB4 and developed in a 4:1 solution 
of water and AZ351B for 10-12~seconds. 
The exposed silicon is etched in a reactive ion etching 
machine with SF$_6$ gas for 2~minutes, followed by 
photoresist removal in 80°C hot PG remover for 30~minutes. 
The sample is again exposed to O$_2$ plasma with 50~W 
power for 2~minutes to remove any residue of photoresist.

\textbf{Step-4 Patterning of the top electrode:}
A 100~nm thick aluminum film is deposited on the samples 
using the sputtering technique. 
Patterning of the top electrode follows the exact same 
procedure as the bottom electrode.

\textbf{Step-5 Etching of the sacrificial layer and release of the drum:}
To avoid any electrostatic effect leading to the collapse 
of the drum, we shorted the two electrodes of the sample 
using a wire bond before the release process. 
The Si-sacrificial layer is etched using a high-pressure 
SF$_6$ etch in a reactive ion system. 
The following are the typical RIE etch parameters: 50W RF power, 
95~mTorr chamber pressure, 100~sccm flow rate 
of SF$_6$, seven cycles of 8~mins of etching process 
with 2~mins break between each cycle.

\section{Measurement setup}

\begin{figure*}
	\begin{center}
		\includegraphics[width = 130 mm]{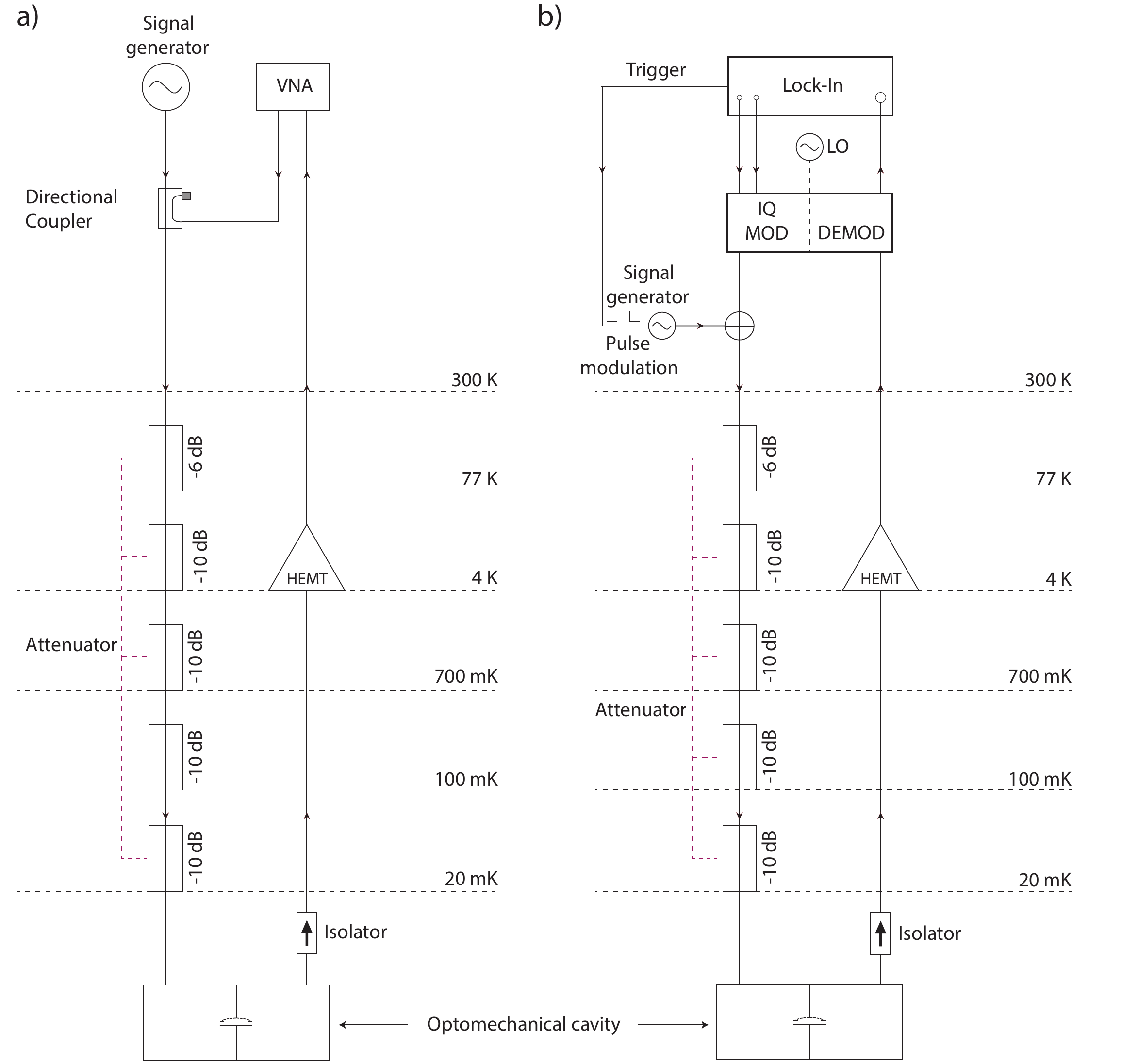}
		\caption{Setup for (a) spectroscopy and (b) time-domain measurements.}
		\label{measurement schemaic}
	\end{center}
\end{figure*}

Fig.~\ref{measurement schemaic}(a) shows the measurement 
set up for the optomechanically induced absorption technique. 
We use a red-detuned strong pump generated by a signal generator 
(R$\&$S - SMF100A)  and a probe tone near the cavity resonance 
frequency provided by a vector network analyser 
(R$\&$S - ZNB20). The time base of both the signal generators 
are synchronized by an external 10~MHz clock. 
We use a directional coupler at the input to combine 
both signals. The input signal goes through a series 
of attenuators at different cooling stages in the dilution 
refrigerator before reaching the input of the cavity. 
The output from the cavity gets amplified by a HEMT amplifier 
at the 4~K stage before reaching the input of the vector network analyser. 

Fig.~\ref{measurement schemaic}(b) shows the time domain 
measurement setup used to perform the optomechanical swap. 
The pump is pulse-modulated using a trigger. The pulse-modulated
pump output is then added to the probe signal.
To maintain vector control over the probe signal, we use a
home-built $IQ$-modulation-demodulation box based on IQ-mixer 
(Marki-IQ4509) for up-conversion, and a Mini-Circuits-ZMX-10G+
three-port mixer for down-conversion. The base-band signals
for the probe are generated from the 2-channel outputs of a 
lock-in amplifier (Zurich instruments - UHFLI). The demodulated
RF output from the demodulator box is sent to the lock-in 
amplifier's inputs to record the quadratures $I(t)$ and $Q(t)$.

\section{Key device parameters}
Table~\ref{my-table} lists some key device parameters
extracted from various experiments.
Calibration of the external coupling rates of the microwave
input and output ports ($\kappa_{e1}$ and $\kappa_{e2}$)
is performed in a separate cooldown where we measured the 
reflection coefficients from both microwave ports using 
two separate HEMT amplifiers.

\begin{table}
	\centering
	\caption{Important device parameters}
	\label{my-table}
	\begin{tabular}{|p{60mm}|p{20mm}|p{15mm}| p{15mm}| } \hline
		Description & Parameter & Value & Unit \\ 
		\hline
		Cavity frequency  & $\omega_c$  & 4.86 & GHz  \\ 
		\hline
		Microwave input coupling rate  & $\kappa_{e1}$  & 90 & kHz  \\ 
		\hline
		Microwave output coupling rate  & $\kappa_{e2}$  & 190 & kHz  \\ 
		\hline
		Cavity linewidth & $\kappa$ & 380 & kHz \\ 
		\hline
		Mechanical frequency & $\omega_m$ & 6.32 & MHz \\
		\hline
		Mechanical dissipation rate & $\gamma_m$ & 20 & Hz \\
		\hline
		Single photon coupling rate & $g_0$ & 165 & Hz \\ 
		\hline
	\end{tabular}
\end{table}

\section{Estimation of single photon coupling rate $g_0$ using simulation}

\begin{figure*}
	\begin{center}
		\includegraphics[width = 85 mm]{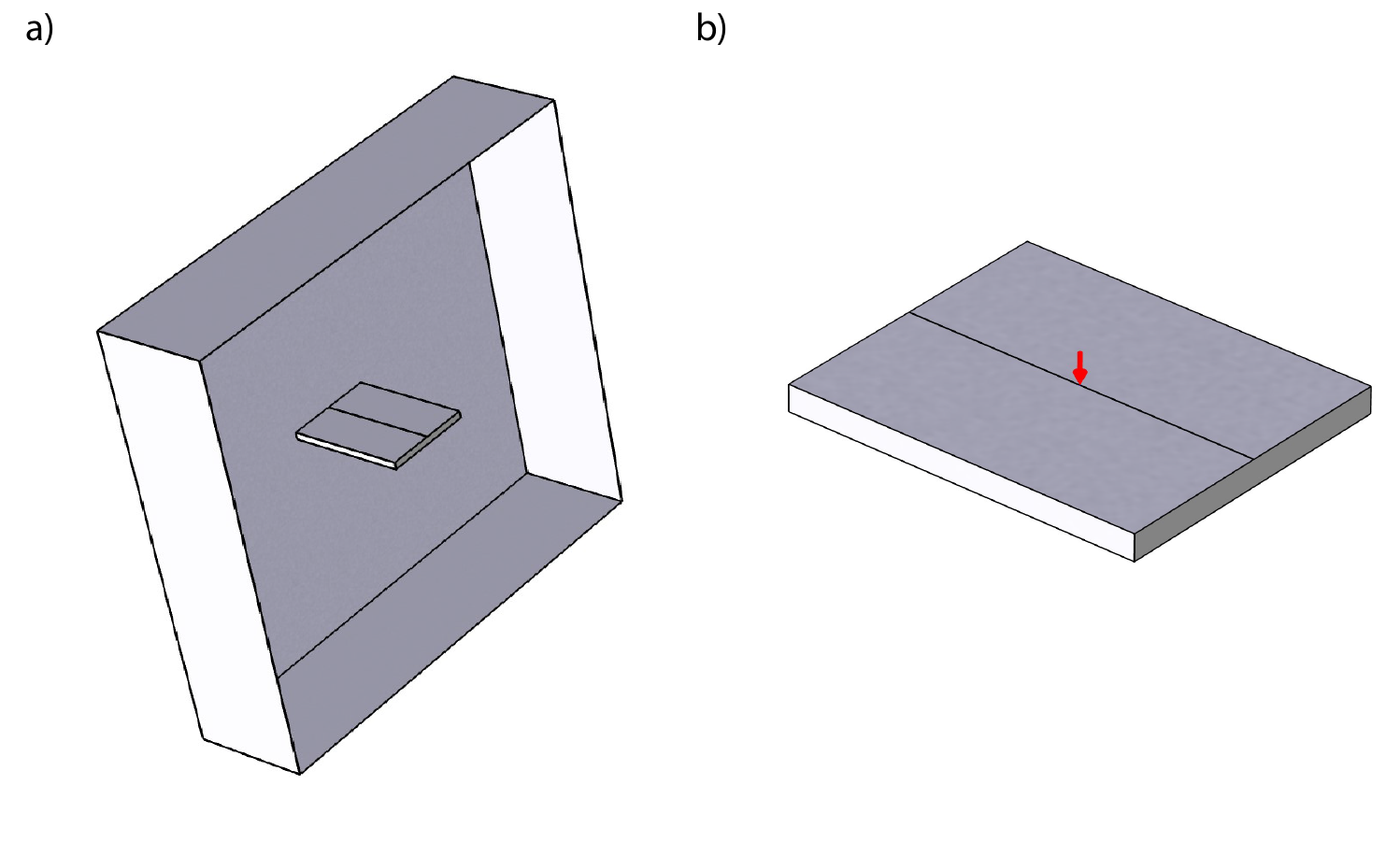}
		\caption{(a) A snippet of the microwave design used in 
			COMSOL showing the 3D cavity and the placement of the 
			sapphire chip. (b) Sapphire chip with two electrodes. The position 
			of the lumped port in place of the capacitor is pointed by the 
			red arrow.}
		\label{Comsol simulation}
	\end{center}
\end{figure*}

We perform electromagnetic simulations using COMSOL to estimate
the single photon coupling strength $g_0$. 
We model a 3D cavity of dimensions $26 \times 26 \times 6$ $\text{mm}^3$ 
and a sapphire chip placed at its center, as shown 
in Fig.~\ref{Comsol simulation}(a). 
The sapphire chip has two electrodes on either side. 
The mechanically-compliant capacitor is modeled as a 
lumped port placed at the center, as shown by the 
red arrow in Fig.~\ref{Comsol simulation}(b). 
We perform a frequency domain simulation by defining the 
Al surfaces as ``perfect electric conductor" boundary 
condition, which then gives us the admittance response of 
the system for the defined frequency range. 
We then add the admittance contribution of the capacitor 
to the admittance response of the simulation in such a 
manner that the zero crossing in the imaginary part of the
admittance curve matches with the experimentally observed 
resonance frequency of the cavity \cite{nigg_black-box_2012}. 
This gives us an estimate of the capacitance of the parallel 
plate capacitor. Thus, we can estimate the gap ($d$) between the 
two plates using the lateral dimensions of the drum. 
The single photon coupling strength for an optomechanical device
can be written as $\eta \omega_c x_{zp}/2d$, where $\eta$ 
is the capacitance participation ratio of the mechanical capacitor
to the total capacitance of the device, $x_{zp}$ is the vacuum 
fluctuation of the mechanical resonator.

\section{Estimation of attenuation in the input line, and calibration of $g_0$}
\begin{figure*}
\begin{center}
\includegraphics[width = 140 mm]{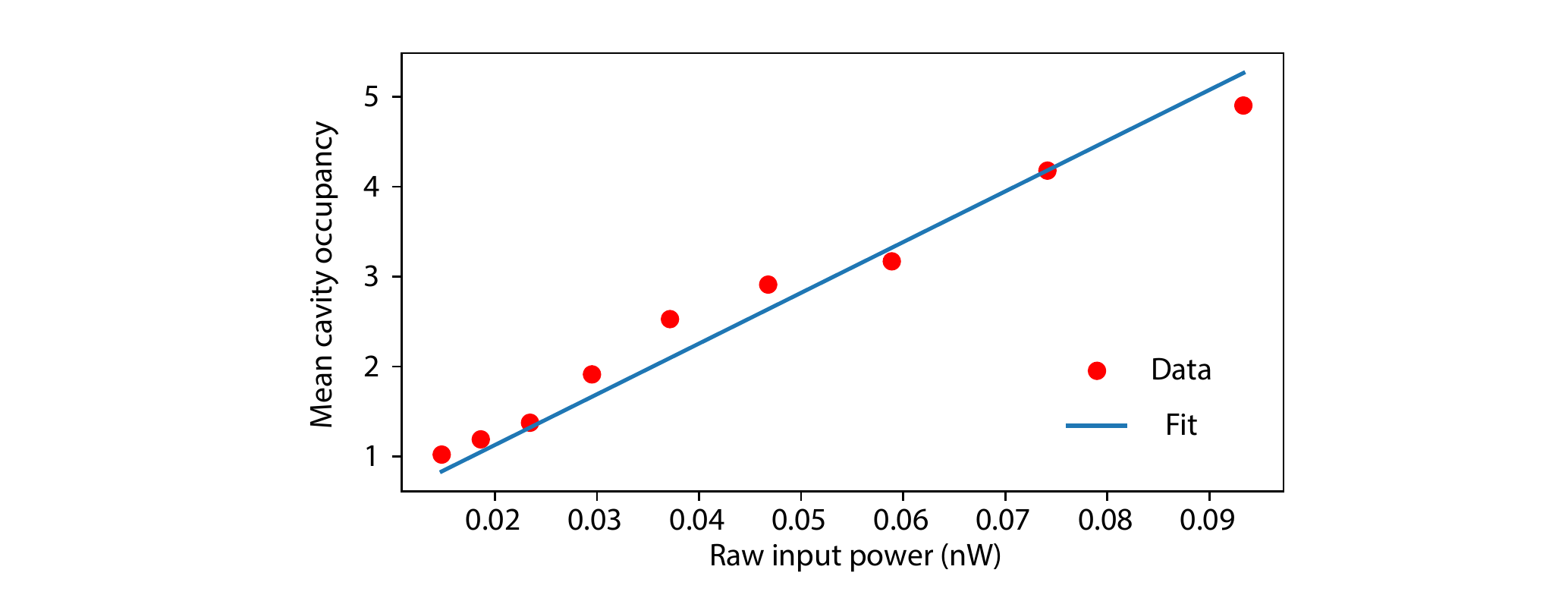}
\caption{The input line attenuation was calibrated by 
calculating the cavity photon population using the 
ac-Stark shift and plotting it against the input 
power at the cable. The fitted parameter was used 
to determine the line attenuation, which was found 
to be 53 dB.}
\label{SI-stark-figure}
\end{center}
\end{figure*}

The input line attenuation was determined in a separate cooldown
using a measurement technique commonly employed in circuit-quantum 
electrodynamics (c-QED) systems to calculate the intra-cavity 
photon population. In this case, a transmon qubit of frequency 
$\omega_q/2\pi=3.6$~GHz was placed inside the 3D cavity.
The frequency shift due to the ac-Stark shift of the transmon 
was measured to determine the mean cavity 
photon population \cite{schuster_ac_2005}.
The frequency of the transmon exhibits a linear relationship 
with the mean cavity photon population given by
$\Tilde{\omega}_q = \omega_q + 2 \chi \Bar{n}$,
where $\Bar{n}$ is mean cavity occupation and $\chi$ is 
the dispersive frequency shift. 
The dispersive shift can be expressed 
as $\chi = J^2 \frac{\alpha}{\Delta (\Delta + \alpha)}$, 
where $\Delta = \omega_q - \omega_c$ is the detuning between 
the qubit and the cavity, $\alpha$ is the qubit anharmonicity, 
and $J$ is the coupling between the qubit and cavity.
The qubit parameters such as $J$ and anharmonicity $\alpha$ 
can be independently measured using single-tone and two-tone
spectroscopy methods. Thus, the calculated dispersive shift 
can be used to calibrate the cavity photon population, and hence
the attenuation of the cables. 
For our device, we find the dispersive shift to be 
$\chi/2\pi = -0.12~\text{MHz}$.
Fig.~\ref{SI-stark-figure} displays the cavity photon population 
as a function of input power at the cable external to the setup. 
By using the total attenuation of the input line as a fitting 
parameter, the attenuation was inferred to be 53~dB for 
the input line.
Combined with the optomechanically induced absorption (OMIA) 
measurements, it further allows us 
to calibrate the single-photon optomechanical coupling 
rate to be $2\pi \times$ 165~Hz, which closely matches with 
the estimation made using COMSOL simulations. 

\section{Estimation of Kerr-shift due to kinetic inductance}

We perform cavity transmission measurements at different 
probe powers. With increased input probe power, 
the cavity shows a shift in the resonant frequency. 
We attribute the shift in the cavity frequency to 
the optomechanical Kerr effect and shift due to the 
kinetic inductance of aluminium film. After removing 
the optomechanical Kerr-shift per photon $-2g^2_0/\omega_m$, 
we attribute the rest of the Kerr-shift to the kinetic 
inductance of the superconducting film. We estimated it to 
be 5~mHz/photon at the highest probe powers used 
in the experiments.

\section{Calculations of the cavity transmission}
The Hamiltonian for an optomechanical cavity system is given by 

\begin{equation}
	\Tilde{\mathcal{H}} = \hbar\omega_c(\hat{x})(\hat{n}+1/2) + \hbar\omega_{m}(b^{\dagger}b+1/2) ,
	\label{final hamiltonian}
\end{equation}
where the operator $b(b^{\dagger})$ is the annihilation(creation)
operator for the mechanical resonator, and 
$\omega_{c} (\omega_{m})$ represents the resonant 
frequency of the microwave cavity (mechanical resonator).
The operator $\hat{n}$ is the photon number operator 
for the microwave cavity. The resonant frequency of the 
microwave cavity depends on the gap between the 
capacitor plates. As the bottom plate is fixed, 
the microwave cavity's resonant frequency depends on the top plate's position
(mechanical oscillator), which is freely suspended. 
When the amplitude of the mechanical oscillator is 
small, we can expand 
$\omega_{c}(\hat{x}) \simeq \omega_c - g_{0}\hat{x}/x_{zp}$,     
where $x_{zp} = \sqrt{\hbar/2m\omega_m}$ is the 
vacuum fluctuations of the mechanical oscillator, $g_{0}$ is the single photon coupling strength, and $m$ is the mass of the mechanical resonator.

With a pump tone at $\omega_{d} = \omega_{c} + \Delta$, 
the cavity gets populated with $n_d$ coherent 
microwave photons with the pump tone frequency. 
The number of photons inside the cavity after 
writing in terms of mean coherent amplitude 
$\sqrt{n_{d}}$ and fluctuating term $\hat{a}$ 
becomes
\begin{equation}
	\hat{n} = (\sqrt{n_d}e^{i\omega_{d}t} + \hat{a}^\dagger)(\sqrt{n_d}e^{-i\omega_{d}t} + \hat{a}) = n_{d} + \sqrt{n_{d}}(\hat{a}^{\dagger}e^{-i\omega_{d}t} + \hat{a}e^{i\omega_{d}t}) + a^{\dagger}a.
\end{equation}

Thus, the resultant Hamiltonian can be written as,
\begin{equation*}
	\begin{aligned}
		\Tilde{\mathcal{H}} = \hbar\omega_{c}n_d + \hbar\omega_c\hat{a}^{\dagger}\hat{a} + \hbar\omega_c\sqrt{n_d}(\hat{a}e^{i\omega_{d}t}+\hat{a}^{\dagger}e^{-i\omega_{d}t})- {\hbar}g_{0}n_{d}( \hat{b}^\dagger+\hat{b}) +\\ {\hbar}g(\hat{a}^{\dagger}e^{-i\omega_{d}t} +  \hat{a}e^{i\omega_{d}t})(\hat{b}^\dagger+\hat{b}) +     \frac{\hbar}{2}g_0(\hat{b}^\dagger + \hat{b}) -{\hbar}g_0(\hat{b}^\dagger+\hat{b}) \hat{a}^\dagger\hat{a} + {\hbar}\omega_m\hat{b^\dagger}\hat{b}.
	\end{aligned}
\end{equation*}
To get the equation of motion for the coupled system, we use the Heisenberg-Langevin approach to get
\begin{equation}
	\Dot{\hat{a}} = -\frac{\kappa}{2}\hat{a} + \sqrt{\kappa_{e1}}\hat{a}_{in} - i\omega_c\sqrt{n_d}e^{-i\omega_dt} + ig_0(\hat{b}^\dagger + \hat{b})\hat{a} + ig(\hat{b}^\dagger + \hat{b})e^{-i\omega_dt} -i\omega_c\hat{a},
\end{equation}
\begin{equation}
	\Dot{\hat{b}} = (-i\omega_m-\frac{\gamma_m}{2})b + ig_0\hat{a}^\dagger\hat{a} + \sqrt{\gamma_m}\hat{b}_{in} + ig_0n_d + i\sqrt{n_d}g_0(\hat{a}e^{-i\omega_dt} + \hat{a}^\dagger e^{-i\omega_dt}),
\end{equation}
where $\kappa_{e1}$ is the external 
coupling rate of the cavity, $\gamma$ is the mechanical 
dissipation rate, and $\hat{a}_{in} (\hat{b}_{in} )$ is the input 
noise operators for the microwave (mechanical) field. 

\par
Taking the Fourier transform of the equation of motions and putting them in a compact matrix form, we get
\begin{equation}
	C(\omega)\begin{bmatrix}
		\hat{a}(\omega + \omega_d)\\
		\hat{b}(\omega)\\
		\hat{a}^\dagger(\omega_d - \omega)\\
		\hat{b}^\dagger(-\omega)\\
	\end{bmatrix} + \begin{bmatrix}
		-\omega_c\sqrt{n_d}\delta(\omega)\\
		g_{0}n_d\delta(\omega)\\
		\omega_c\sqrt{n_d}\delta(\omega)\\
		-g_{0}n_{d}\delta(\omega)\\
	\end{bmatrix} = \begin{bmatrix}
		i\sqrt{\kappa_{e1}}\hat{a}_{in}(\omega + \omega_d)\\
		i\sqrt{\gamma_m}\hat{b}_{in}(\omega)\\
		i\sqrt{\kappa_{e1}}\hat{a}^{\dagger}_{in}(\omega_d - \omega)\\
		i\sqrt{\gamma_m}\hat{b}^{\dagger}_{in}(-\omega)\\ 
	\end{bmatrix}.
\end{equation}

The mode-coupling matrix $C(\omega)$ is given by
\begin{equation}
	C(\omega) = \begin{bmatrix}
		{\chi_a}^{-1}(\omega + \omega_d) & g & 0 & g \\
		g & {\chi_b}^{-1}(\omega) & g & 0\\
		0 & -g & -{\chi_a}^{-1}(\omega_d - \omega)^* & -g\\
		-g & 0 & -g & -{\chi_b}^{-1}(-\omega)^*\\
	\end{bmatrix},
\end{equation}
where
\begin{equation}
	{\chi_a}(\omega) = 1/(\omega - \omega_c + i\kappa/2),
\end{equation}
\begin{equation}
	{\chi_b}(\omega) = 1/(\omega - \omega_m + i\gamma_m/2)
\end{equation}
are the complex susceptibility functions for the cavity and mechanical mode, respectively.

We use optomechanically induced absorption (OMIA) to 
couple the cavity and mechanical motion and get the 
resultant response of the dressed mode. 
Here, we provide a strong red-detuned pump at a 
frequency $\omega_{c}-\omega_{m}$ and a 
weak probe tone near cavity resonance frequency, 
$\omega_c$. The resultant transmission through 
the cavity at a frequency $\omega$ is then given by 

\begin{equation}
	T(\omega) = i\sqrt{\kappa_{e1}\kappa_{e2}}{\chi^{eff}_{a}(\omega)},
\end{equation}
where ${\chi^{eff}_a}(\omega)$ is given by $(C^{-1}(\omega))_{11}$, $\kappa_{e1}(\kappa_{e2})$ is the input(output) coupling rate.

\section{Linear Stability test (Routh-Hurwitz stability criteria)}

To understand the instabilities at large pump powers, 
we consider the optomechanical Kerr-nonlinearity and 
the kinetic inductance of aluminium film.
We start with the full optomechanical Hamiltonian in 
the rotating frame of the pump frequency. It can be written as,
\begin{equation}
	\mathcal{H} = - \Delta \hat{a}^\dagger\hat{a} - \frac{\alpha_c}{2}\hat{a}^\dagger\hat{a}^\dagger\hat{a}\hat{a} + \omega_m \hat{b}^\dagger\hat{b} - g_0 \hat{a}^\dagger\hat{a}(\hat{b} + \hat{b}^\dagger) + iE(\hat{a} + \hat{a}^\dagger),
	\label{nonlinear hamiltonian}
\end{equation}
where $\alpha_c$ is the Kerr-coefficient to model 
the kinetic inductance. Using the Heisenberg-Langevin 
approach and the semi-classical approximation, the equations of motions 
can be written as,
\begin{equation}
	\dot{\alpha} = ({i}\Delta - \frac{\kappa}{2})\alpha + {i}\alpha_c \left| \alpha \right|^2 \alpha + {i}g_0 \alpha (\beta + \beta^{\ast}) + E,
\end{equation}
\begin{equation}
	\dot{\beta} = -({i}\omega_m + \frac{\gamma_m}{2})\beta + {i}g_0 \left| \alpha \right|^2,
\end{equation}
where $E =  \sqrt{\kappa_{e1}}~\hat{a}_{in}$, 
$\alpha = \langle \hat{a} \rangle$ and 
$\beta = \langle \hat{b} \rangle$. 
By representing $\alpha$ and $\beta$ in the complex form, 
we get
\begin{equation}
	\alpha = x + {i}y,
\end{equation}
\begin{equation}
	\beta = p + {i}q.
\end{equation}
Therefore,
\begin{equation}
	\dot{x} = f_1(x, y, p, q) = -\frac{\kappa}{2} x - \Delta y - 2g_0 p y  -\alpha_c y (x^2 + y^2) +E
	\label{equation fo real alpha},
\end{equation}
\begin{equation}
	\dot{y} = f_2(x, y, p, q) =  \Delta x - \frac{\kappa}{2} y + 2g_0 p x + \alpha_c x (x^2 + y^2)
	\label{equation fo imag alpha},
\end{equation}
\begin{equation}
	\dot{p} = f_3(x, y, p, q) = - \frac{\gamma_m}{2} p + \omega_m q
	\label{equation fo real beta},
\end{equation}
\begin{equation}
	\dot{q} = f_4(x, y, p, q) = g_0 x^2 + g_0 y^2 -\omega_m p - \frac{\gamma_m}{2} q
	\label{equation fo imag beta}.
\end{equation}
Following the approach used in Ref.~\cite{bakemeier_route_2015,roque_nonlinear_2020}, we 
obtain the fixed points of the system by setting 
the first derivative of the real and imaginary parts of 
both $\alpha$ and $\beta$ to zero, \textit{i.e.}
$\dot{x} = \dot{y} = \dot{p} = \dot{q} = 0$.
Combining Eq.~\ref{equation fo real alpha} and 
Eq.~\ref{equation fo imag alpha} and representing 
$x,y,q$ in terms of $p$, we get
\begin{equation}
	x = \frac{\kappa}{2} \frac{E}{A^2 + (\frac{\kappa}{2})^2},
\end{equation}
\begin{equation}
	y = \frac{A~E}{A^2 + (\frac{\kappa}{2})^2},
\end{equation}
\begin{equation}
	q = \frac{\gamma_m}{2 \omega_m} p,
	\label{q in terms of p}
\end{equation}
where
\begin{equation}
	A = \Delta + B~p
\end{equation}
and
\begin{equation}
	B = \left [ 2g_0 +\frac{\alpha_c}{g_0}(\omega_m + \frac{\gamma_m^2}{4\omega_m} ) \right ].
\end{equation}
Substituting the above calculated values of $x,y,q$ in Eq. ~\ref{equation fo imag beta}, we get a cubic polynomial equation, 
\begin{equation}
	B^2\left(\omega_m + \frac{\gamma^2_m}{4\omega_m} \right) p^3 + 2B\Delta\left(\omega_m + \frac{\gamma^2_m}{4\omega_m} \right)~p^2 + \left(\Delta^2 + \frac{\kappa^2}{4}\right)\left(\omega_m + \frac{\gamma^2_m}{4\omega_m} \right)~p - g_0 E^2 = 0.
\end{equation}
The roots of the polynomial will give the fixed point 
solutions for $x$, $y$, $q$. 
Considering the roots 
to be $(\Bar{x}, \Bar{y}, \Bar{p}, \Bar{q}$),
the nature of the fixed points can be understood by 
considering the time evolution of a small 
perturbation around these points.
We define a small perturbation around the fixed points
as $Z_i = k_i - \Bar{k}_i$,
where $i = 1,2,3,4$ corresponding to 
$(k_1,k_2,k_3,k_4) \equiv (x,y,p,q)$.

To obtain the time evolution of $Z_i$, we solve
$\dot{Z}_i = \dot{k}_i = f_i(x, y, p, q)$ by expanding
the $f_i$ polynomials around the fixed point and 
retaining the first-order terms as shown below:

\begin{equation}
	f_j(x, y, p, q) \approx f_j\Big|_{x,y,p,q = \Bar{x}, \Bar{y}, \Bar{p}, \Bar{q}} +\sum_{k=x,y,p,q} (k_i - \Bar{k}_i)\frac{\partial f_j}{\partial k_i}\Big|_{x,y,p,q = \Bar{x}, \Bar{y}, \Bar{p}, \Bar{q}}.
\end{equation}

This allows us to obtain the four coupled equations of motion
for the small perturbations around fixed points as,
\begin{equation}
	\frac{d}{dt}
	\begin{bmatrix}
		Z_1 \\
		Z_2 \\
		Z_3 \\
		Z_4
	\end{bmatrix} = 
	\begin{bmatrix}
		\frac{\partial f_1}{\partial x} &  \frac{\partial f_1}{\partial y} &  \frac{\partial f_1}{\partial p} &  \frac{\partial f_1}{\partial q} \\ 
		\frac{\partial f_2}{\partial x} &  \frac{\partial f_2}{\partial y} &  \frac{\partial f_2}{\partial p} &  \frac{\partial f_2}{\partial q} \\
		\frac{\partial f_3}{\partial x} &  \frac{\partial f_3}{\partial y} &  \frac{\partial f_3}{\partial p} &  \frac{\partial f_3}{\partial q} \\
		\frac{\partial f_4}{\partial x} &  \frac{\partial f_4}{\partial y} &  \frac{\partial f_4}{\partial p} &  \frac{\partial f_4}{\partial q}
	\end{bmatrix} \begin{bmatrix}
		Z_1 \\
		Z_2 \\
		Z_3 \\
		Z_4
	\end{bmatrix}.
\end{equation}
Upon substituting the values of $f_i$'s, and 
evaluating the derivative at the fixed points,
we get
\begin{equation}
	\frac{d}{dt}
	\begin{bmatrix}
		Z_1 \\
		Z_2 \\
		Z_3 \\
		Z_4
	\end{bmatrix} = S
	 \begin{bmatrix}
		Z_1 \\
		Z_2 \\
		Z_3 \\
		Z_4
	\end{bmatrix},
\end{equation}
where the evolution matrix $S$ is given by 
\begin{equation}
	S = 
	\begin{bmatrix}
		- \frac{\kappa}{2} -2\alpha_c \Bar{x} \Bar{y} & -\Delta - 2 g_0 \Bar{p} - \alpha_c \Bar{x}^2 - 3\alpha_c \Bar{y}^2 & - 2 g_0 \Bar{y} & 0 \\ 
		\Delta + 2 g_0 \Bar{p} + 3\alpha_c \Bar{x}^2 + \alpha_c \Bar{y}^2 & - \frac{\kappa}{2} + 2\alpha_c \Bar{x} \Bar{y} & 2 g_0 \Bar{x} & 0 \\
		0 & 0 & - \frac{\gamma_m}{2} & \omega_m \\
		2 g_0 \Bar{x} & 2 g_0 \Bar{y} & -\omega_m & - \frac{\gamma_m}{2}
	\end{bmatrix}
\end{equation}

The solution of the matrix equation will 
have the following form,
\begin{equation}
	Z(t) = \sum_{i=1}^{4}c_iW_i \exp{\lambda_i t},
	\label{sols}
\end{equation}
where $c_i$'s are the constants of integration, $W_i$'s and  
$\lambda_i$' are the eigenvectors and eigenvalues of 
the matrix $S$.

From Eq.~\ref{sols}, it is evident that stable solutions
are only possible iff all the eigenvalues have 
a negative real part.
To cross-check the results from the nonlinear
optomechanically coupling can be obtained
in a straightforward manner by setting 
$\alpha_c$ to zero, as given by,
\begin{equation}
	\frac{d}{dt}
	\begin{bmatrix}
		Z_1 \\
		Z_2 \\
		Z_3 \\
		Z_4
	\end{bmatrix} = 
	\begin{bmatrix}
		- \frac{\kappa}{2} & -\Delta - 2 g_0 \Bar{p} & - 2 g_0 \Bar{y} & 0 \\ 
		\Delta + 2 g_0 \Bar{p} & - \frac{\kappa}{2}& 2 g_0 \Bar{x} & 0 \\
		0 & 0 & - \frac{\gamma_m}{2} & \omega_m \\
		2 g_0 \Bar{x} & 2 g_0 \Bar{y} & -\omega_m & - \frac{\gamma_m}{2}
	\end{bmatrix} \begin{bmatrix}
		Z_1 \\
		Z_2 \\
		Z_3 \\
		Z_4
	\end{bmatrix}.
\end{equation}

Instability boundaries obtained from the above formalism
by choosing $\alpha_c/2\pi=$~0, 5, 12.5~mHz/photo are 
included in Fig.~4(b) of the main text.

\section{Measurement of instability boundary}

\begin{figure*}
\centering
\includegraphics[width = 85 mm]{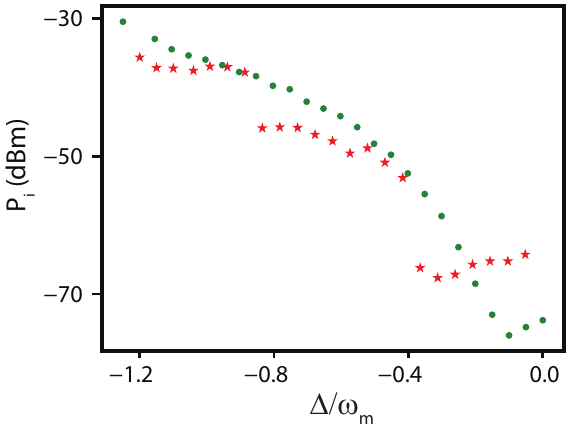}
\caption{The green dots represent the measured 
instability boundary when the pump power is swept upwards. 
The red stars represent the instability boundary 
obtained while sweeping the pump power downwards.}
\label{instability_hyster}
\end{figure*}

The measurement of the instability boundary is 
performed in two ways. In the first method, we 
send a pump signal and measure the power spectral 
density (PSD) of the output signal near the pump frequency.
The pump power is manually changed while sweeping it
upwards and the threshold power at the onset of instability
is recorded. 
Subsequently, the power is reduced to the start value,
and the measurement for the next detuning is carried
out in the same way.
In the second method, we divide the parameter space 
spanned by $\Delta$ and $P_i$ into a regular grid. The resolution of drive frequency and pump power is~300~kHz and 0.1~dBm respectively.
We use an automated data recording which records 
the power spectral density for different parameter 
values on the grid. 
The instability boundary, period-doubling/tripling 
and chaos are then extracted by analyzing the records of 
PSD.
We do not see any substantial difference in the manual
and automated datasets. We, however observe a hysteresis
when the pump power is swept upwards or downwards as shown
in Fig.~\ref{instability_hyster}. 
We suspect that this hysteresis might come from
the heating of the device or from the duffing behavior of 
the mechanical resonator.

\section{From optomechanically-induced absorption to the ultrastrong coupling}
\begin{figure*}
\centering
\includegraphics[width = 90 mm]{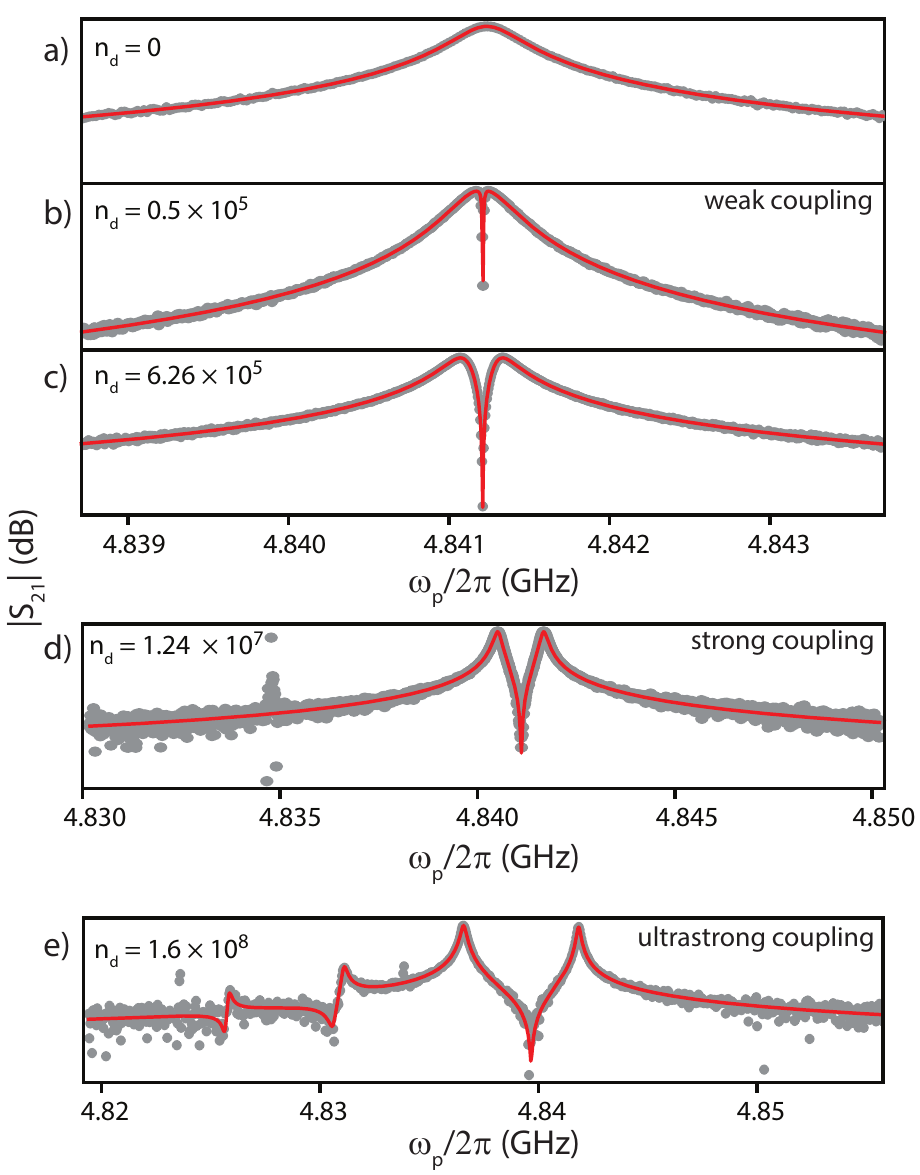}
\caption{(a) shows the cavity transmission $|S_{21}|$ 
at the base temperature without any pump. 
(b)-(e) shows the OMIA measurement as pump strength 
is increased. The gray circles represent the 
measurement data, and the red curve represents the theoretical fit.}
\label{USC}
\end{figure*}

To characterize the ultrastrong coupling limit, 
we start with a weak pump signal at the red sideband. 
We gradually increase the pump power as we move from 
the weak coupling limit to the strong coupling limit. 
At high pump powers, the pump frequency is adjusted 
as the cavity resonance frequency shifts down due to the 
static Kerr-shift. Fig.~\ref{USC} shows the transmission 
spectra of the cavity at different pump photons.

\section{Theoretical model and measurement of OMIA in the time domain}

To theoretically model the observed behavior of Fig.~3(d)
in the main text, 
we start with the Hamiltonian in a rotating frame 
at the drive frequency as given by \ref{nonlinear hamiltonian}.
By including the probe part, the total Hamiltonian can 
be written as, 
\begin{equation}
\mathcal{H} \approx - \Tilde{\Delta} \hat{a}^\dagger\hat{a} + \omega_m \hat{b}^\dagger\hat{b} + g (\hat{a} +\hat{a}^\dagger)(\hat{b} + \hat{b}^\dagger) + S_p (\hat{a}\mathrm{e}^{{i}\Omega t} + \hat{a}^\dagger \mathrm{e}^{-{i}\Omega t}),
\end{equation}
where $\Tilde{\Delta} = \Delta + \frac{2 g_0^2 \alpha^2}{\omega_m}$, $g = g_0 \alpha$ and $\Omega = \omega_p - \omega_d$.
From the equation above, we can obtain the following equation of motions (EOMs):
\begin{subequations}
	\begin{equation}
		\langle\dot{\hat{a}}\rangle = ({i}\Tilde{\Delta} - \frac{\kappa}{2})\langle\hat{a}\rangle + (-{i}g)\langle\hat{b}\rangle + (-{i}g)\langle\hat{b}^\dagger\rangle + S^\ast(t),
	\end{equation}
	\begin{equation}
		\langle\dot{\hat{a}}^\dagger\rangle = (-{i}\Tilde{\Delta} - \frac{\kappa}{2})\langle\hat{a}^\dagger\rangle + ({i}g)\langle\hat{b}\rangle + ({i}g)\langle\hat{b}^\dagger\rangle + S(t),
	\end{equation}
	\begin{equation}
		\langle\dot{\hat{b}}\rangle = (-{i}g)\langle\hat{a}\rangle + (-{i}g)\langle\hat{a}^\dagger\rangle + (-{i}\omega_m - \frac{\gamma}{2})\langle\hat{b}\rangle,
	\end{equation}
	\begin{equation}
		\langle\dot{\hat{b}}^\dagger\rangle = ({i}g)\langle\hat{a}\rangle + ({i}g)\langle\hat{a}^\dagger\rangle + ({i}\omega_m - \frac{\gamma}{2})\langle\hat{b}^\dagger\rangle,
	\end{equation}
\end{subequations}
where $S(t) = {i}S_p \mathrm{e}^{{i}\Omega t}$. 
Similar to the approach described earlier, these
coupled equations can be solved in the frequency domain. 
The solution obtained can then be transformed back
to the time-domain using the inverse Fourier transform. 
The solution of the EOM in the frequency domain can be written in a compact form as,
\begin{equation}
	X[\omega] = \mathcal{B}(\omega)~r[\omega],
	\label{fourier EOM}
\end{equation}
where 
\begin{equation}
	\mathcal{B}(\omega) = \begin{bmatrix}
		-{i}\omega-{i}\Tilde{\Delta} + \frac{\kappa}{2} & 0 & {i}g & {i}g \\
		0 & -{i}\omega+{i}\Tilde{\Delta} + \frac{\kappa}{2} & -{i}g & -{i}g \\
		{i}g & {i}g & -{i}\omega+{i}\omega_m + \frac{\gamma}{2} & 0 \\
		-{i}g & -{i}g & 0 & -{i}\omega-{i}\omega_m + \frac{\gamma}{2}
	\end{bmatrix}^{\scalebox{1}{-1}},
\end{equation}
\begin{equation}
	X(\omega) = \begin{bmatrix}
		\langle\hat{a}\rangle(\omega) \\ \langle\hat{a}^\dagger\rangle(\omega) \\ \langle\hat{b}\rangle(\omega) \\ \langle\hat{b}^\dagger\rangle(\omega)
	\end{bmatrix},
\quad
\text{ and }
\quad
	r(\omega)
	= \begin{bmatrix}
		- 2\pi{i}S_p  \delta(\omega - \Omega) \\
		2\pi{i}S_p  \delta(\omega + \Omega)\\
		0 \\
		0
	\end{bmatrix}.
\end{equation}

After performing the inverse Fourier transformation
on the eq.~\ref{fourier EOM}, we can write the total 
solution as,

\begin{equation}
	\begin{bmatrix}
		\langle\hat{a}\rangle(t) \\ \langle\hat{a}^\dagger\rangle(t) \\ \langle\hat{b}\rangle(t) \\ \langle\hat{b}^\dagger\rangle(t)
	\end{bmatrix} = \begin{bmatrix}
		X_1 & X_2 & X_3 & X_4
	\end{bmatrix}
	\begin{bmatrix}
		a_1\mathrm{e}^{\lambda_1 t} \\
		a_2\mathrm{e}^{\lambda_2 t} \\
		a_3\mathrm{e}^{\lambda_3 t} \\
		a_4\mathrm{e}^{\lambda_4 t}
	\end{bmatrix} + 
	{i}S_p \begin{bmatrix}
		-\mathcal{B}_{11}(\Omega) \mathrm{e}^{-{i}\Omega t} + \mathcal{B}_{12}(-\Omega) \mathrm{e}^{{i}\Omega t} \\ 
		-\mathcal{B}_{21}(\Omega) \mathrm{e}^{-{i}\Omega t} + \mathcal{B}_{22}(-\Omega) \mathrm{e}^{{i}\Omega t} \\ 
		-\mathcal{B}_{31}(\Omega) \mathrm{e}^{-{i}\Omega t} + \mathcal{B}_{32}(-\Omega) \mathrm{e}^{{i}\Omega t} \\ 
		-\mathcal{B}_{41}(\Omega) \mathrm{e}^{-{i}\Omega t} + \mathcal{B}_{42}(-\Omega) \mathrm{e}^{{i}\Omega t}
	\end{bmatrix}
\end{equation}
Here, the $a_i$'s are integration constants. The $\lambda_i$'s and $X_i$'s are eigenvalues and eigenvectors 
of the matrix $M$ defined as,
\begin{equation}
	M = \begin{bmatrix}
		{i}\Tilde{\Delta} - \frac{\kappa}{2} & 0 & -{i}g & -{i}g \\
		0 & -{i}\Tilde{\Delta} - \frac{\kappa}{2} & {i}g & {i}g \\
		-{i}g & -{i}g & -{i}\omega_m - \frac{\gamma}{2} & 0 \\
		{i}g & {i}g & 0 & {i}\omega_m - \frac{\gamma}{2}
	\end{bmatrix}.
	\label{matrix-A}
\end{equation}
 
 If the initial boundary condition is defined at some point in time $t_0$, then the integration constant can be calculated as,
\begin{equation}
	\begin{bmatrix}
		a_1\mathrm{e}^{\lambda_1 t_0} \\
		a_2\mathrm{e}^{\lambda_2 t_0} \\
		a_3\mathrm{e}^{\lambda_3 t_0} \\
		a_4\mathrm{e}^{\lambda_4 t_0}
	\end{bmatrix} = 
	\begin{bmatrix}
		X_1 & X_2 & X_3 & X_4
	\end{bmatrix}^{\scalebox{1}{-1}}
	\begin{bmatrix}
		\langle\hat{a}\rangle(t_0) + {i}S_p\mathcal{B}_{11}(\Omega)\mathrm{e}^{-{i}\Omega t_0}  - {i}S_p\mathcal{B}_{12}(-\Omega)\mathrm{e}^{{i}\Omega t_0} \\ \langle\hat{a}^\dagger\rangle(t_0)  + {i}S_p\mathcal{B}_{21}(\Omega)\mathrm{e}^{-{i}\Omega t_0}  - {i}S_p\mathcal{B}_{22}(-\Omega)\mathrm{e}^{{i}\Omega t_0}\\ \langle\hat{b}\rangle(t_0) + {i}S_p\mathcal{B}_{31}(\Omega)\mathrm{e}^{-{i}\Omega t_0}  - {i}S_p\mathcal{B}_{32}(-\Omega)\mathrm{e}^{{i}\Omega t_0}\\ \langle\hat{b}^\dagger\rangle(t_0) + {i}S_p\mathcal{B}_{41}(\Omega)\mathrm{e}^{-{i}\Omega t_0}  - {i}S_p\mathcal{B}_{42}(-\Omega)\mathrm{e}^{{i}\Omega t_0}.
	\end{bmatrix}
\end{equation}

\begin{figure*}
\begin{center}
\includegraphics[width = 140 mm]{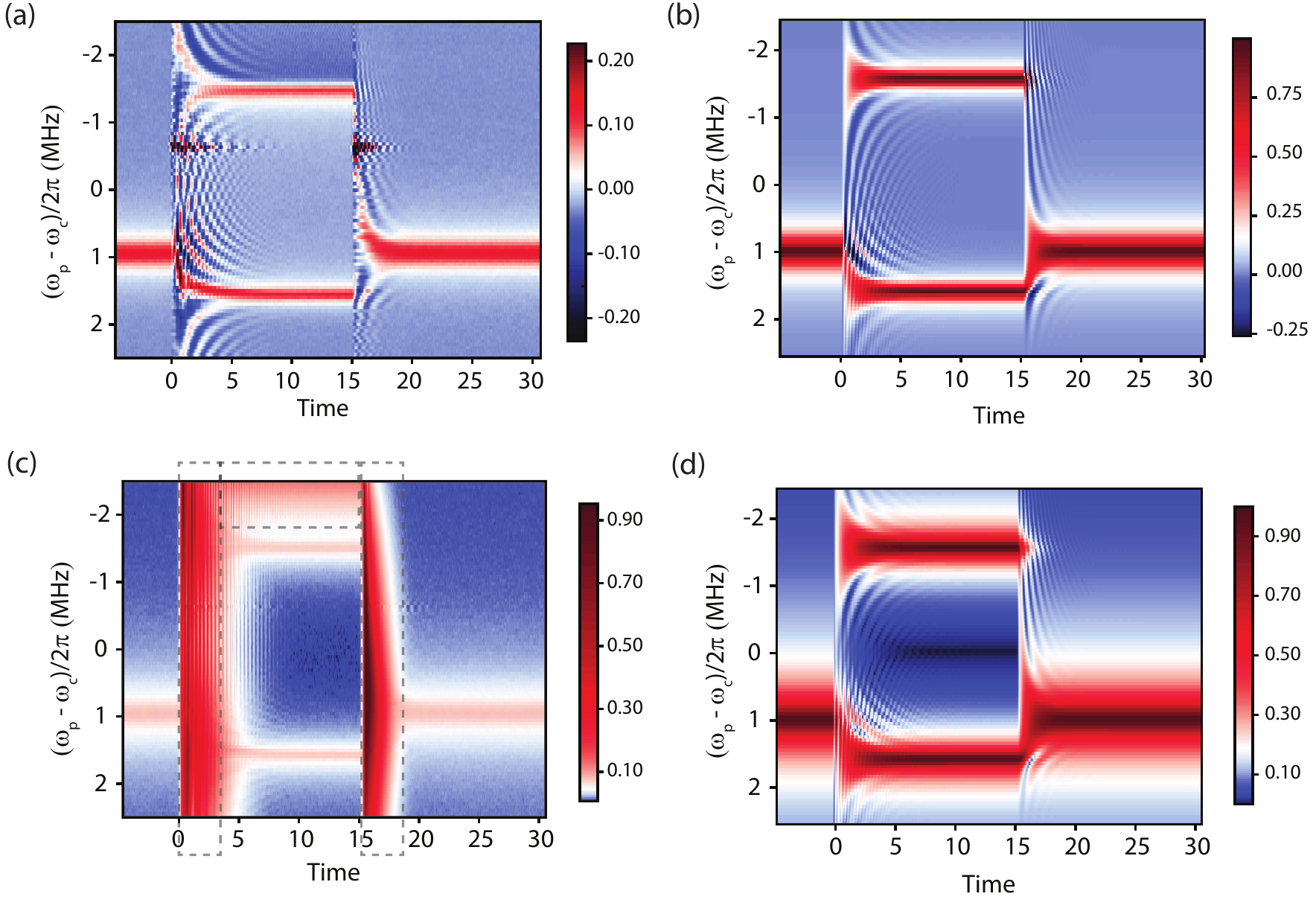}
\caption{(a) shows the measurement of the imaginary 
part of the probe signal. (b) shows the 
corresponding results from the theoretical calculations. 
(c) shows the measurement of the absolute value of the 
probe signal. The dotted box shows the presence of an
excess offset signal coming from the pulsed pump signal
captured by the lock-in demodulator. 
(d) show the theoretical calculations results for the
magnitude of probe signal.}
\label{Time domain OMIA}
\end{center}
\end{figure*}

Since this solution is obtained in a frame rotating at the
pump frequency, to obtain the solution at the probe 
frequency we shift the frame by $\Omega$, and neglect the 
terms rotating at $2\Omega$.
The solution described above are plotted in Fig.~\ref{Time domain OMIA}(b,d).
We emphasize here that the homogeneous solution 
containing $\mathrm{e}^{\lambda t}$ terms is
responsible for the fringing pattern when the 
pump is turned on. The in-homogeneous solution gives 
the steady-state solution.

For measurement of the optomechanically-induced 
absorption in the time-domain, we send a weak probe 
signal near the cavity resonant frequency and a 
pulsed pump signal of duration $\approx$15~$\micro$s 
near the red sideband of the cavity.
The frequency of the probe signal is swept across 
the cavity frequency and the time domain response 
for each probe frequency is recorded. 
Fig.~\ref{Time domain OMIA}(a) shows the measurement 
of the imaginary part of the probe signal from the cavity.
Panel (b) shows the plot of the solution from the theoretical
calculations as described above, which show a quite decent
match between experimental and theoretical results.

We would like to emphasize the technical challenge 
of such measurements. In order to capture the oscillatory
nature of the transient solution, one must demodulate the 
microwave signal using a sufficiently large demodulation
bandwidth (small time constant) of the lock-in amplifier. 
At the same time, the large demodulation bandwidth must not
capture the pump signal which leaks through the cavity.
Since the oscillation frequency of the transient solution is 
order $2g$, and pump-probe detuning is order $\omega_m$, it 
makes such measurements technically difficult as $2g$ approaches
$\omega_m$. Such measurement artifacts are more evident if we 
plot the magnitude ($R$) of the probe signal, as shown in 
Fig.~\ref{Time domain OMIA}(c). The large value of magnitude
below $\Omega/2\pi=-2$~MHz for $0\le t\le15~\mu$s, is due to 
the pump signal falling into the demodulator bandwidth.
In addition, this pump leakage also plagues the measurement
when the pump is either turned on or turned off.
At the switching event, the pump signal spreads in 
the frequency domain and spills into the bandwidth 
of the demodulator. It results in an ``excess" value 
in $R$, which decays over a time-scale
set by the cavity-dissipation rate and the demodulator bandwidth.
Results of the magnitude calculations are shown in Fig.~\ref{Time domain OMIA}(d).

\section{Zoomed-in plot of parametric instability}

\begin{figure*}
\centering
\includegraphics[width = 85 mm]{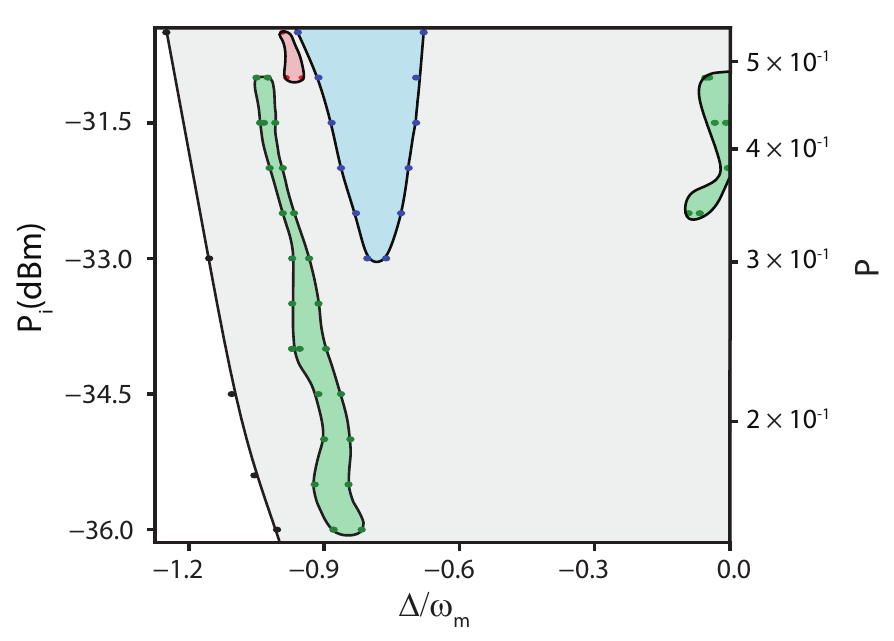}
\caption{Zoomed-in plot of Fig.4(b) of the main text.}
\label{zoomed instability}
\end{figure*}

Fig.~\ref{zoomed instability} shows the detailed view of 
the parametric instability region showing the regions of 
period-doubling, period-tripling, and chaotic behaviors.

\newpage


%